\documentclass[10pt,twocolumn,letterpaper]{article}

\usepackage[pagenumbers]{cvpr} % To force page numbers, e.g. for an arXiv version

\usepackage[dvipsnames]{xcolor}

\definecolor{cvprblue}{rgb}{0.21,0.49,0.74}
\usepackage[pagebackref,breaklinks,colorlinks,citecolor=cvprblue]{hyperref}

\usepackage{algorithm}
\usepackage{algorithmic}
\usepackage{graphicx}
\usepackage{amsmath}
\usepackage{amssymb}
\usepackage{booktabs}
\usepackage{multirow}

\usepackage{times}
\usepackage{epsfig}

\def\paperID{*****} % *** Enter the Paper ID here
\def\confName{3DV\xspace}
\def\confYear{2025\xspace}

\title{Deformable 3D Shape Diffusion Model}
\author{Dengsheng Chen \quad Jie Hu \quad Xiaoming Wei\\
Meituan\\
{\tt\small \{chendengsheng,hujie39,weixiaoming\}@meituan.com}
\and
Enhua Wu\\
SKLCS, Institute of Software, Chinese Academy of Sciences\\
{\tt\small weh@ios.ac.cn}
}
\begin{document}
\maketitle
\begin{abstract}
The Gaussian diffusion model, initially designed for image generation, has recently been adapted for 3D point cloud generation. However, these adaptations have not fully considered the intrinsic geometric characteristics of 3D shapes, thereby constraining the diffusion model's potential for 3D shape manipulation. To address this limitation, we introduce a novel deformable 3D shape diffusion model that facilitates comprehensive 3D shape manipulation, including point cloud generation, mesh deformation, and facial animation.
Our approach innovatively incorporates a differential deformation kernel, which deconstructs the generation of geometric structures into successive non-rigid deformation stages. By leveraging a probabilistic diffusion model to simulate this step-by-step process, our method provides a versatile and efficient solution for a wide range of applications, spanning from graphics rendering to facial expression animation.
Empirical evidence highlights the effectiveness of our approach, demonstrating state-of-the-art performance in point cloud generation and competitive results in mesh deformation. Additionally, extensive visual demonstrations reveal the significant potential of our approach for practical applications. Our method presents a unique pathway for advancing 3D shape manipulation and unlocking new opportunities in the realm of virtual reality.
\end{abstract}
\section{Introduction}

\begin{figure*}
    \centering
    \includegraphics[width=0.9\linewidth]{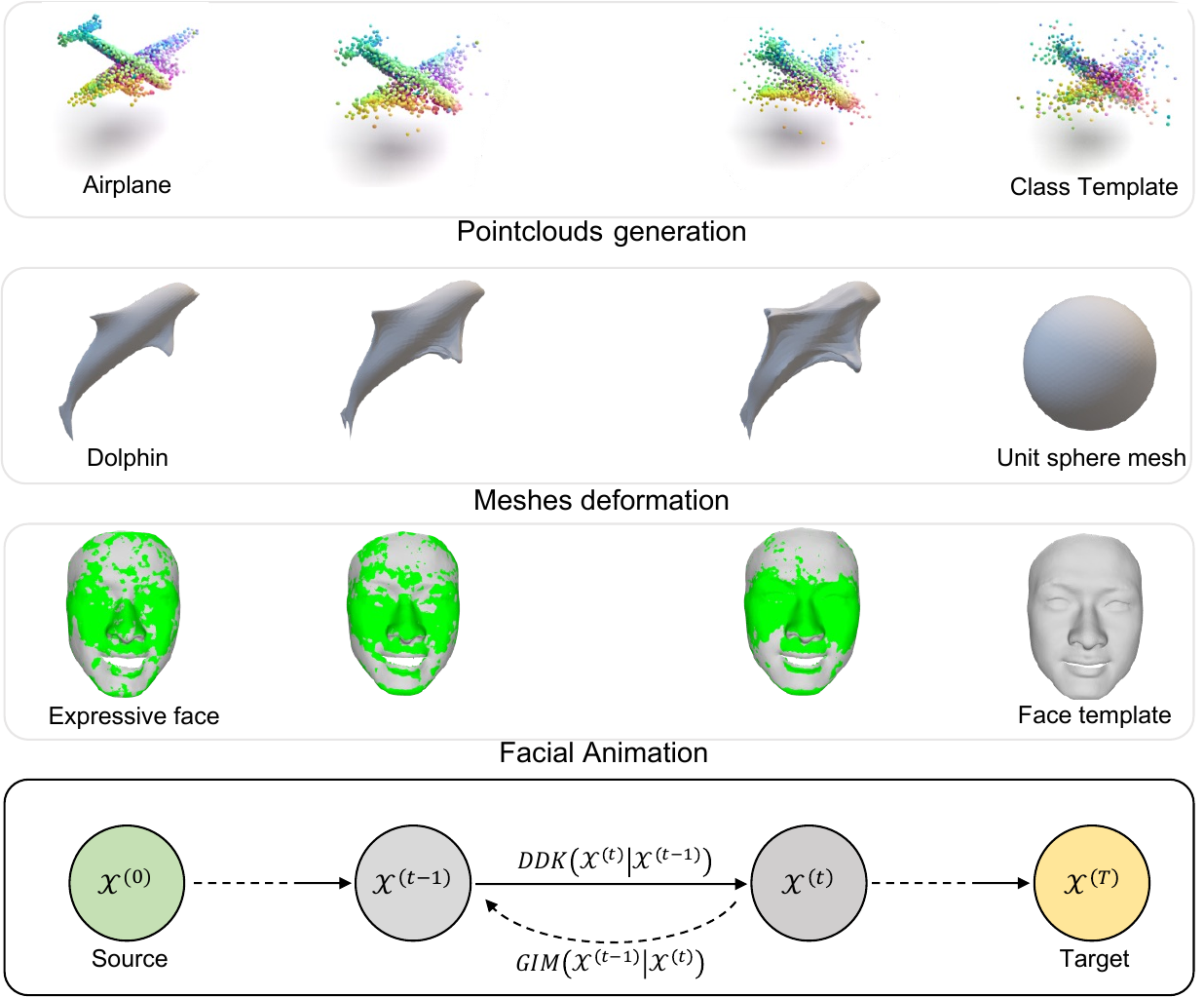}
    \caption{Illustration of the geometric imitation procedure (4th row).
   First, we diffuse a given source shape $\mathcal{X}^{(0)}$ to a well-initialized template shape $\mathcal{X}^{(T)}$ using DDK in a sequence $\{\mathcal{X}^{(0)}, \mathcal{X}^{(1)}, \cdots \mathcal{X}^{(T)} \}$. Subsequently, we use the DDM to reverse the diffusion process and obtain the desired shape.
   Our proposed method is versatile and can be applied to various tasks such as point cloud generation (1st row), mesh deformation (2nd row), and facial animation (3rd row). 
    }
    \label{fig: teaser}
\end{figure*}

The task of developing a generative model for three-dimensional (3D) shapes, which includes point clouds and meshes, has become a pivotal challenge with numerous applications~\cite{suchde2022point,xu2021voxel,guo2020deep}. The Gaussian diffusion model has demonstrated exceptional performance in image generation tasks. Building on previous studies~\cite{zeng2022lion,luo2021diffusion}, our research aims to extend the diffusion model to support the generation of 3D shapes.

To gain a comprehensive understanding of 3D shape generation as a diffusion process, we interpret the discrete coordinates that comprise the shape as particles within a non-equilibrium thermodynamic system. This system interacts with a heat bath, causing the particle positions to evolve stochastically. Over time, these particles undergo diffusion, gradually dispersing throughout the 3D space—a phenomenon known as the diffusion process. Concurrently, noise is added at each time step to progressively transform the initial particle distribution into a simple noise distribution~\cite{luo2021diffusion}.

\begin{figure}[t]
    \centering
    \includegraphics[width=\linewidth]{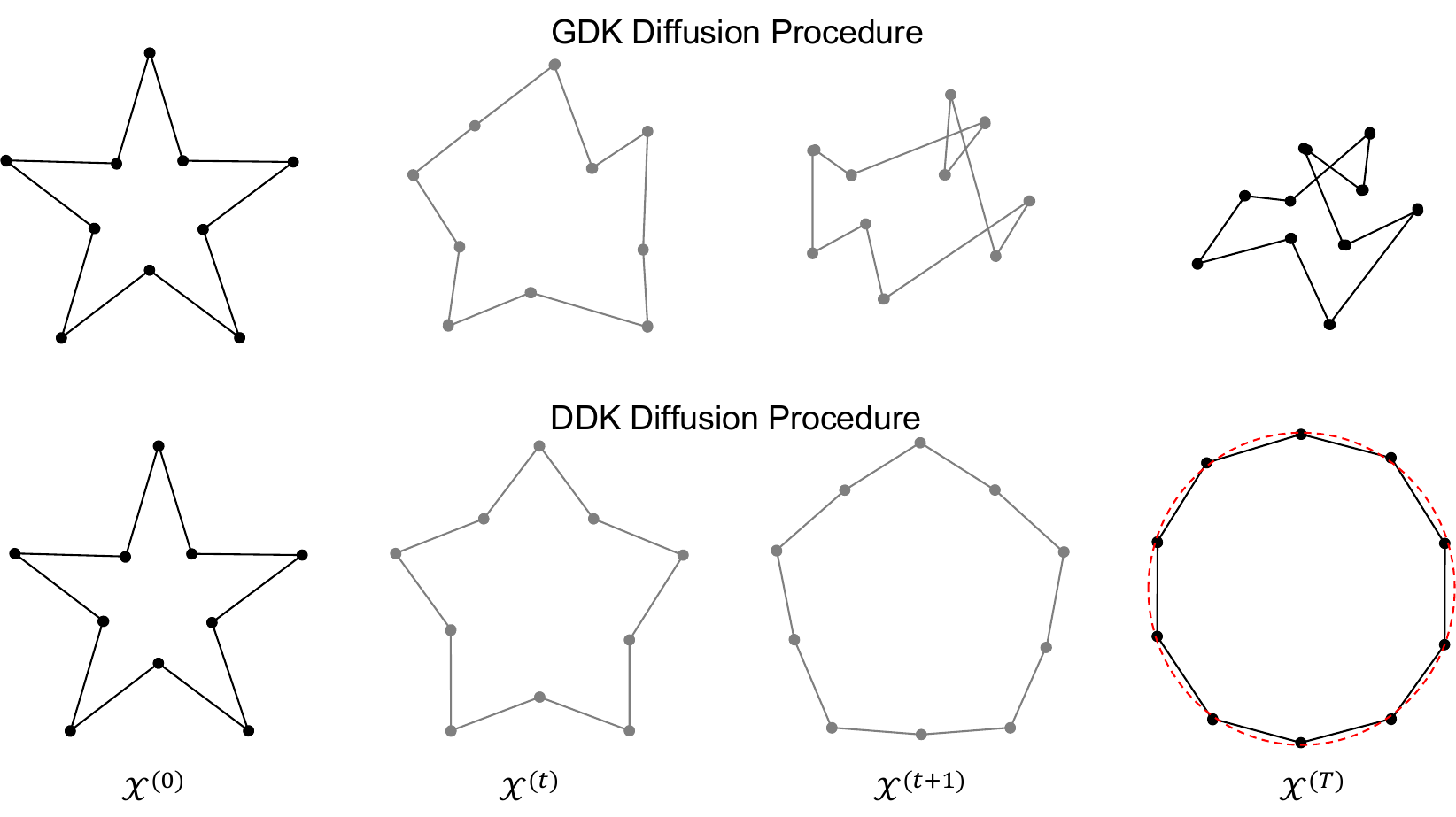}
    \caption{
Comparison of diffusion procedures guided by GDK and DDK. 
As depicted in the first row, using only GDK introduces Gaussian noise to vertices at each step, which quickly erodes the geometric structure of the shape, resulting in disordered point clouds where the edges intersect, thereby disrupting the topology of the meshes. 
In contrast, the proposed DDK can deform shapes based on multiple shape regularizations, producing well-structured manifold meshes. 
    }
    \label{fig: ddk-vs-gdk}
\end{figure}

Drawing an analogy, we can link the distribution of coordinates in point clouds or meshes to a noise distribution through the diffusion process. This concept is fundamental to the probabilistic diffusion model that underpins our methodology for 3D shape generation. However, unlike pixel data, 3D data is characterized by the interplay between spatial (coordinate) positions and geometric feature information. Consequently, the introduction of noise to the coordinates not only alters their spatial positions but also disrupts the local geometric structure. Therefore, the traditional diffusion process becomes significantly more challenging to regulate when applied to 3D data, causing meaningful geometric information to dissipate rapidly within a few steps, as illustrated in Fig.~\ref{fig: ddk-vs-gdk}. This constraint presents substantial hurdles in modeling fine-grained geometric deformations and limits the efficacy of mesh generation techniques.

To address the challenges associated with 3D data manipulation, we introduce a novel deformable 3D shape diffusion model. Our model employs a unique Differential Deformation Kernel (DDK) to diffuse a geometric distribution into a predefined template distribution, in contrast to the commonly utilized Gaussian Diffusion Kernel (GDK) in image data. By decomposing the intricate geometric structure into numerous subtle, consecutive samples via the DDK, we aim to capture the gradual deformation characteristics of geometric structures, leveraging the power of the imitation process. This diffusion process enables us to manage 3D data in a geometrically-aware manner.

To facilitate fine-grained geometric deformation and restore the original geometric distribution from a given template distribution, we introduce a methodology that inversely simulates the diffusion process. Unlike image reverse diffusion, which primarily models the posterior distribution of data, our approach directly regresses the final timestep structure based on the current one using a neural network. This stepwise regression process allows us to incrementally reconstruct the original geometric structure.

Our approach represents a significant breakthrough in the development of mesh generation techniques for complex 3D models. Our contributions are summarized as follows:

\begin{itemize}
\item We propose a novel geometric imitation model for 3D shape manipulation, grounded in the diffusion process of non-equilibrium thermodynamics.
\item Our model demonstrates state-of-the-art performance in point cloud generation and competitive performance in mesh deformation, as evidenced by experimental results.
\item Extensive experiments and visualizations highlight the substantial application potential of our method in various fields, including point cloud generation, mesh deformation, graphics rendering, and animation production.
\end{itemize}

\section{Geometric Imitation Models}
\begin{algorithm}[t]
    \caption{Differential deformation kernel~(DDK).}
    \label{alg: differential deformation}
    \begin{algorithmic}[1]
        \REQUIRE A well-initialized~(target) shape $\mathcal{X}^{(T)}$ and source shape $\mathcal{X}^{(0)}$.
        \FOR{\texttt{$t=1 \to T$}}
        	\STATE $\mathbf{o}_i^{(t-1\to t)} \gets - \frac{\partial \mathcal{L}(\mathcal{X}^{(t-1)}, \mathcal{X}^{(0)})}{\mathbf{x}_i^{(t-1)}}$.
			\STATE $\mathbf{x}_{i}^{(t)} = \mathbf{x}_{i}^{(t-1)} + \eta \mathbf{o}_{i}^{(t-1\to t)} + \mathcal{ N }(\mathbf{0}, \beta_t \mathbf{I})$.
		\ENDFOR
        \RETURN $\{\mathcal{X}^{(0)}, \mathcal{X}^{(1)}, \cdots, \mathcal{X}^{(T)}\}$.
    \end{algorithmic}
\end{algorithm}
This section delves into the geometric imitation model we have developed for the forward and reverse diffusion processes of 3D shapes. We also discuss relevant shape regularizations pertinent to our model. 

Subsequently, we articulate our proposed differential deformation kernel, which facilitates the diffusion of samples within a geometry-conscious reDDMe. Comprehensive descriptions of the training and sampling algorithms we employ are provided. 

\subsection{Formulation of Diffusion Process}
A point cloud or mesh $\mathcal{X}$ can be depicted as a set of $n$ vertices $\{ \mathbf{x}_i \}_{i=1}^n$ along with associated normals $\{ 
\mathbf{n}_{i} \} _{i=1}^n$ and $e$ edges $\{ \mathbf{e}_i \}_{i=1}^e$.

We perceive the set of particles $\mathcal{X}^{(0)}$ as the initial state of a dynamic thermodynamic system. With the progression of time, the \textit{diffusion process} gradually disperses the vertices into a disordered set of points, culminating in a noise distribution. The forward diffusion process is conceptualized as a Markov chain~\cite{jarzynski1997equilibrium}:
\begin{equation}
    q(\mathcal{X}^{(1:T)}|\mathcal{X}^{(0)}) = \prod_{t=1}^T q(\mathcal{X}^{(t)} | \mathcal{X}^{(t-1)}),
\end{equation}
where $q(\mathcal{X}^{(t)} | \mathcal{X}^{(t-1)})$ is the Markov diffusion kernel. 

Prior research~\cite{sohl2015deep,luo2021diffusion} has adopted the strategy of considering each vertex $\mathbf{x}_i$ as being sampled \textit{independently} from a point distribution $q(\mathbf{x}_i^{(0)})$. These studies employ a Gaussian diffusion kernel~(GDK) that progressively introduces Gaussian noise to the vertices, and the posterior distribution is modeled to reverse this process.

Given that 3D data encapsulates both spatial~(coordinate) position and feature~(geometry) information, introducing noise solely to the coordinates not only displaces the spatial position but also disrupts the local geometric structure. As a result, the Gaussian diffusion kernel is not suitable for diffusing 3D data that adheres to a non-rigid deformation in a well-manifold manner, and it fails to maintain the fine-grained geometric structure. To circumvent this issue and facilitate the diffusion of samples in a geometric-aware reDDMe that guides the model to sample from a geometry distribution rather than a point distribution, we propose a novel Markov diffusion kernel, termed the Differential Deformation Kernel~(DDK), denoted by $q(\mathcal{X}^{(t)}| \mathcal{X}^{(t-1)}, \mathcal{X}^{(0)}) \sim \mathcal{D} (\mathcal{X}^{(t)}; \mathcal{X}^{(t-1)}, \mathcal{X}^{(0)}, \beta_t \mathbf{I})$, which will be further elaborated in Sec.~\ref{sec:ddk}.
Considering that the normal vector $\mathbf{n}_i^{(t)}$ can be readily recomputed from the neighboring vertex $\mathcal{N}_{\mathbf{x}_i^{(t)}}$, and that the edge connection $\mathbf{e}_{i}^{(t)}$ persists unaltered throughout the diffusion process, our primary objective is to synthesize a novel set of vertices $\{\mathbf{x}_i^{(t)}\}_{i=1}^{n}$ for $\mathcal{X}^{(t)}$ that demonstrates a cohesive geometric structure. 

To achieve this, we envision the generation process as the inverse of the diffusion process. Specifically, we initiate by sampling vertices from a rudimentary noise distribution $p(\mathbf{x}_i^{(T)})$, and subsequently guide the sampled vertices through the inverse Markov chain, parameterized by an approximate estimation $p_\theta (\mathbf{x}_i^{(t-1)} | \mathbf{x}_i^{(t)})$, until the target distribution $p(\mathbf{x}_i^{(0)})$ is formed.

We express the inverse process for generation as:
\begin{align}
    p _\theta (\mathcal{X} ^{(0:T)}) &:= p(\mathcal{X}^{(T)}) \prod _{t=1} ^T p_{\mathbf{\theta}} (\mathcal{X}^{(t-1)} | \mathcal{X}^{(t)}), \\
    p _{\mathbf{\theta}} (\mathcal{X} ^{(t-1)} | \mathcal{X}^{(t)}) &:= \mathcal{F_\theta} (\mathcal{X}^{t-1}; \mathcal{X}^{(t)}, \beta_t \mathbf{I}) \label{eq:posterior}
\end{align}
Here, $\mathcal{F_\theta}$ denotes an estimated posterior distribution parameterized by $\mathbf{\theta}$. We refer to this inverse process as \textit{Geometric imitation learning}, which will be elaborated in Sec.~\ref{sec:DDM}.
The initial distribution $p(\mathcal{X}^{(T)})$ can be sampled from a standard normal distribution, a unit sphere, or a specific template shape, contingent on the task at hand. Hence, $p(\mathcal{X}^{(T)})$ is a constant that is independent of $\theta$.

\subsection{Shape regularization}
\begin{algorithm}[htb]
    \caption{Deformable imitation learning.}
    \label{alg: imitation learning}
    \begin{algorithmic}[1]
    	\REQUIRE A well-initialized shape $\mathcal{X}^{(T)}$, training dataset $\mathbb{S}$.
        \REPEAT
            \STATE $\mathcal{X}^{(0)} \sim \mathbb{S}$
            \STATE $\mathcal{X}^{(0:T)} = \{\mathcal{X}^{(0)}, \mathcal{X}^{(1)}, \cdots\, \mathcal{X}^{(T)}\}$ $\gets$ DDK
            \STATE Sample from $\mathcal{X}^{(0:T)}$ and take gradient descent on 
            $$\nabla _{\theta} \parallel \varphi_{\theta}(\mathbf{x}^{(t-1)})- \mathbf{x}^{(t)} \parallel _2^2$$
        \UNTIL{converged}
    \end{algorithmic}
\end{algorithm}

Prior to introducing the aforementioned DDK, we first delineate some shape regularizations employed during non-rigid transformation.
Unless otherwise specified, we utilize $\mathbf{p}$ and $\mathbf{q}$ to denote a vertex in point cloud (or mesh) $\mathcal{P}$ and $\mathcal{Q}$, respectively, until the conclusion of this sub-section.
Given that these constraints are frequently employed in related literature, we provide a succinct description of the impact of each constraint. For a more comprehensive understanding, please refer to ~\cite{gupta2020neural}.
\paragraph{Chamfer Distance Loss}
Prior to discussing the previously mentioned DDK, we initially establish some shape regularizations utilized in non-rigid transformations. Unless otherwise specified, we employ $\mathbf{p}$ and $\mathbf{q}$ to denote a vertex in pointcloud (or mesh) $\mathcal{P}$ and $\mathcal{Q}$ respectively, throughout this subsection. The Chamfer distance loss, which is often used to measure the distance between two point clouds when the corresponding point relationship is undefined, is defined as:
\begin{equation}
    \ell _{c} (\mathcal{P}, \mathcal{Q}) = \sum _{\mathbf{p} \in \mathcal{P}} \min _{ \mathbf{q} \in \mathcal{Q}} \parallel \mathbf{p} - \mathbf{q} \parallel _2^2 + \sum _{ \mathbf{q} \in\mathcal{Q}} \min_{\mathbf{p}\in \mathcal{P}} \parallel \mathbf{p} - \mathbf{q} \parallel _2^2.
\end{equation}
While effective in driving vertices towards their accurate positions, this loss alone is insufficient for generating a high-quality 3D mesh, as the optimization can easily fall into local minima. The network may produce extreme deformations to favor local consistency, which can be particularly detrimental when the estimated point cloud significantly deviates from the ground truth, resulting in the creation of flying vertices. To address these issues, we suggest the inclusion of additional shape regularizations to capture more intricate information about the underlying geometry.

\paragraph{Normal Consistency Regularization}
The normal consistency regularization, a surface normal loss capturing high-order properties, can be expressed as follows:
\begin{equation}
    \ell _{n}(\mathcal{P}) = \sum_{\mathbf{p} \in \mathcal{P}} \sum_{\mathbf{k}\in \mathcal{N}_{\mathbf{p}}} \parallel \langle  \mathbf{p} -\mathbf{k}, \mathbf{n}_{\mathbf{p}} \rangle  \parallel _2^2
\end{equation}
where $\mathbf{q}$ is the closest vertex to $\mathbf{p}$ identified during the Chamfer distance loss calculation, $\mathbf{k}$ is the neighboring pixel of $\mathbf{p}$, $\langle \cdot ,\cdot \rangle $ denotes the inner product of two vectors, and $\mathbf{n}_{\mathbf{p}}$ is the observed surface normal.

This loss aims to ensure that the edge connecting a vertex with its neighbors remains perpendicular to the observation from the neighboring points. Although this loss may not necessarily be zero except for a planar surface, its optimization can effectively enforce consistency between the normal of a locally fitted tangent plane and the observed surface normal. This approach has proven beneficial in our experiments. Additionally, this normal loss function is fully differentiable, simplifying its optimization.

\paragraph{Laplacian Regularization}
The Laplacian term, crucial in preserving local details, ensures that adjacent vertices move in the same direction. We first define a Laplacian coordinate for each vertex $\mathbf{p}$ as: 
\begin{equation}
    \delta _{\mathbf{p}} = \mathbf{p} - \sum_{\mathbf{k}\in \mathcal{N}(\mathbf{p})} \frac{\mathbf{k}}{\parallel \mathcal{N}(\mathbf{p}) \parallel},
\end{equation}
and the Laplacian regularization is defined as: 
\begin{equation}
    \ell_{l}(\mathcal{P}) = \sum_{\mathbf{p} \in \mathcal{P}} \parallel \delta_{ \mathbf{p}}^\prime - \delta_{\mathbf{p}} \parallel _2^2,
\end{equation}
where $\delta _{\mathbf{p}}$ and  $\delta _{\mathbf{p}}^\prime $ are the Laplacian coordinates of a vertex before and after a deformation block, respectively.
The Laplacian term restricts excessive vertex movement. By promoting the collective movement of neighboring vertices, this term assists in preserving the local mesh structure and preventing self-intersections.
\paragraph{Edge Length Regularization.}

To mitigate the problem of flying vertices, which may result in elongated edges within the mesh, we propose an edge length regularization strategy. This regularization penalty is mathematically expressed as:
\begin{equation}
    \ell _{ e } (\mathcal{P}) = \sum_{\mathbf{p} \in\mathcal{P}}\sum_{\mathbf{k}\in \mathcal{N}(\mathbf{p})} \parallel \mathbf{p} - \mathbf{k} \parallel _2^2.
\end{equation}
It is important to note that this edge length regularization is exclusively applied to neighboring vertices that are connected via edges. 

\paragraph{Potential Energy Regularization.}

To tackle the problem of vertex clustering, which can result in an uneven distribution of vertices, we introduce a potential energy regularization. This can be mathematically formulated as:
\begin{equation}
    \ell_{p}(\mathcal{P}) = \sum_{ \mathbf{p} \in \mathcal{P}}\sum_{ \mathbf{k} \in \mathcal{N}_{\mathbf{p}}} \frac{1}{1 + \parallel \mathbf{p} - \mathbf{k} \parallel_2^2}.
\end{equation}
This regularization differs from the edge length regularization as it not only imposes penalties on vertex clustering to regulate the point distribution, but it also influences all neighboring vertices, irrespective of whether they are interconnected by an edge. 
This regularization encourages vertices to maintain an optimal distance from each other, ensuring the overall potential energy of the point cloud or mesh remains relatively low, thereby fostering a uniform vertex distribution across the entire mesh.
\subsection{Differential Deformation Kernel (DDK)}\label{sec:ddk}

To facilitate sample diffusion in a geometrically cognizant manner, we introduce a novel Markov diffusion kernel, termed the Differential Deformation Kernel (DDK):
\begin{equation}
    q(\mathcal{X}^{(t)} | \mathcal{X}^{(t-1)}, \mathcal{X}^{(0)}) := \mathcal{D} (\mathcal{X}^{(t)}; \mathcal{X}^{(t-1)}, \mathcal{X}^{(0)}, \beta_t \mathbf{I}) 
\end{equation}
Here, the variance schedule hyper-parameters $\beta_1, \cdots, \beta_T$ dictate the quantum of noise introduced at each timestep.

Employing an initial source shape $\mathcal{X}^{(0)}$, we leverage shape regularizations in conjunction with a back-propagation algorithm to execute non-rigid deformation, thereby progressively attaining the target noised shape $\mathcal{X}^{(T)}$. The implementation is elucidated in Algorithm~\ref{alg: differential deformation}.

We can formally define the shape regularizations of this gradient-based DDK as follows:
\begin{equation}
\mathcal{L}(\mathcal{X}^{(t)}, \mathcal{X}^{(0)}) = \underbrace{
\lambda^{T} \cdot \ell(\mathcal{X}^{(t)}) 
}_{\text{shape regularizations}} + \underbrace{\lambda _{c} \ell_{c} (\mathcal{X}^{(t)}, \mathcal{X}^{(0)})}_{\text{supervised loss}}
    \label{eq: loss function of differential deformation}
\end{equation}
Here, $\ell(\mathcal{X}^{(t)})$ = $[\ell_{n} (\mathcal{X}^{(t)}), \ell_{l} (\mathcal{X}^{(t)}), \ell_{e}(\mathcal{X}^{(t)}),  \ell_{p}(\mathcal{X}^{(t)})]$, where $\lambda = [\lambda_n, \lambda_l, \lambda_e, \lambda_p]$ is a hyper-parameter that modulates the influence of distinct shape regularizations. $\lambda_c$ is a hyper-parameter that modulates the influence of supervised Chamfer distance.

By invoking the chain rule of differentiation, we can compute the offset $\mathbf{o}_{i}^{(t-1\to t)}$ for each vertex $\mathbf{x}_{i}^{(t-1)}$ as follows:
\begin{equation}
    \mathbf{o}_{i}^{(t-1 \to t)} = - \frac{\partial \mathcal{L}(\mathcal{X}^{(t-1)}, \mathcal{X}^{(0)})}{\mathbf{x}_{i}^{(t-1)}}.
\end{equation}
Consequently, the forward diffusion process facilitated by DDK can be expressed as:
\begin{equation}
    \mathbf{x}_{i}^{(t)} = \mathbf{x}_{i}^{(t-1)} + \eta \mathbf{o}_{i}^{(t-1\to t)} + \mathcal{ N }(\mathbf{0}, \beta_t \mathbf{I}).
\end{equation}
where $\mathcal{ N }(\mathbf{0}, \beta_t \mathbf{I})$ represents a Gaussian distribution with zero mean and $\beta_t\mathbf{I}$ variance and $\eta$ denotes the step size. As is evident, DDK is an approximate Gaussian kernel function that takes into account the plausible geometric structure of the noisy 3D space while incorporating Gaussian noise. A comparative visualization of GDK and DDK is presented in Fig.~\ref{fig: ddk-vs-gdk}.
\subsection{Deformable Imitation Learning} \label{sec:DDM}

In order to invert the diffusion process, it is necessary to estimate the posterior distribution $p (\mathcal{X} ^{(t-1)} | \mathcal{X}^{(t)})$ as delineated in Equation~\ref{eq:posterior}. However, instead of approximating the posterior distribution, we introduce a neural network with a learnable parameter $\theta$, termed as the \textit{Deformable Diffusion Model~(DDM)}, that directly regresses $\mathcal{X}^{(t-1)}$ from $\mathcal{X}^{(t)}$. 

The aim of training the inverse diffusion process is to enable the model to emulate the non-rigid deformation process as follows:
\begin{equation}
    L(\theta) = \sum_{i=1}^{n} \parallel \varphi_\theta(\mathbf{x}_{i}^{(t-1)}) - \mathbf{x}_{i}^{(t)} \parallel_2^2.
\end{equation}
The point-to-point training objective $L$ can be optimized using standard backpropagation algorithms. The learning process is elaborated in Algorithm~\ref{alg: imitation learning}.

\subsection{3D Shape Sampling}
\begin{algorithm}[htb]
    \caption{Sampling for generation.}
    \label{alg: generation}
    \begin{algorithmic}[1]
    	\REQUIRE A well-initialed template shape $\mathcal{X}^{(T)}$.
        \FOR{$ t = T, \cdots, 1 $}
            \STATE $\epsilon \sim \mathcal{ N }(\mathbf{0}, \beta_t \mathbf{I})$
            \STATE $\mathbf{x}^{(t-1)} \gets \varphi_{\theta}(\mathbf{x}^{(t)} +  \epsilon)$
        \ENDFOR
        \RETURN $\mathcal{X}^{(0)}$
    \end{algorithmic}
\end{algorithm}

$\mathcal{X}^{(T)}$ is a provided template, which could either be a randomly sampled noise point cloud or a well-structured sphere mesh. 
With $\mathcal{X}^{(T)}$, we can iteratively sample a high-quality 3D shape $\mathcal{X}^{(0)}$ from a learned geometric distribution model $\varphi_{\theta}$, as illustrated in Algorithm~\ref{alg: generation}. This enables our algorithm to be implemented in a wider range of scenarios.

%\subsection{Optimized Initialization of $\protect\mathcal{X}^{(T)}$}
\subsection{Optimized Initialization of $X^{(T)}$}

Our experimental results indicate that an optimized initialization of $\mathcal{X}^{(T)}$ can significantly accelerate the diffusion steps required for DDK to converge to $\mathcal{X}^{(T)}$. Consequently, we propose two strategies for selecting an appropriate $\mathcal{X}^{(T)}$ for both point clouds and meshes.

\paragraph{Data-driven Initialization of $\mathcal{X}^{(T)}$ for Point Clouds.}

The diffusion model exhibits a robust expressive capability, enabling the diffusion of $\mathcal{X}^{(0)}$ to a randomly initialized $\mathcal{X}^{(T)}$ directly. Furthermore, the reverse diffusion process can restore $\mathcal{X}^{(0)}$ within several hundred steps, as evidenced in \cite{luo2021diffusion}. Nevertheless, an optimized initialization of $\mathcal{X}^{(T)}$ can further reduce the steps required for DDK to converge to $\mathcal{X}^{(T)}$, particularly when generating high-fidelity point clouds with intricate structures.

To derive the data-driven $\mathcal{X}^{(T)}$, we iteratively apply DDK to various $\mathcal{X}^{(0)}$ over several thousand steps. This procedure results in the formation of an \textit{average shape} for the training data, which expedites the imitation learning process, especially for 3D shapes within a single class.

\paragraph{Template-based Initialization of $\mathcal{X}^{(T)}$ for Meshes.}

When working with meshes, it is essential to preserve the topologies among faces. However, ensuring this for a randomly initialized $\mathcal{X}^{(T)}$ can be challenging. Data-driven methods are not suitable for generating an average shape as point clouds due to the significant variations in topologies among different shapes.

For specific cases such as facial expression animation, we utilize the provided base facial shape (the one without any expression) as $\mathcal{X}^{(T)}$. For general mesh generation tasks, we employ a unit sphere mesh as $\mathcal{X}^{(T)}$. Although it may require more steps for DDK to deform the unit sphere mesh to $\mathcal{X}^{(0)}$, it ensures the production of a well-manifold mesh.

\subsection{Equispaced Sampling}

Existing methods~\cite{song2020denoising} can expedite the sampling process under GDK due to the favorable properties of Gaussian distribution. Additionally, we propose an equispaced sampling trick from the geometric perspective to accelerate the reverse diffusion process for DDK.

Owing to the advantages of DDK in preserving the well-manifold structure, we can employ an equispaced sampling method to generate a subsequence $\mathcal{X}_{s} = \{\mathcal{X}^{(T)}, \mathcal{X}^{(T-i)}, \mathcal{X}^{(T-2i)}, \cdots\, \mathcal{X}^{(0)}\}$ from $\{\mathcal{X}^{(T)}, \mathcal{X}^{(T-1)}, \cdots, \mathcal{X}^{(0)}\}$, where $i$ represents the timestep interval between two consecutive samples.
\section{Experimental Evaluation}

This section elucidates the superiority of the Deformable Diffusion Model~(DDM) in generating point clouds and meshes through three distinct tasks. 

Given that our methods directly sample a shape from $\mathcal{X}^{(T)}$ without the guidance of a latent vector, conventional metrics for evaluating the quality of point clouds or meshes are inapplicable. Consequently, we primarily assess the quality of various algorithms by visualizing the shapes they generate.

\begin{table*}[htb]
\centering
\resizebox{\linewidth}{!}{
\begin{tabular}{|c|c|c|c|c|c|c|c|c|}
\hline
Model & Object Type & COV~(\%,$\uparrow$) CD & COV~(\%,$\uparrow$) EMD & MMD~($\downarrow$) CD & MMD~($\downarrow$) EMD & 1-NNA~(\%, $\downarrow$) CD & 1-NNA~(\%, $\downarrow$) EMD & JSD~($\downarrow$) \\ \hline
PC-GAN~\citeyear{achlioptas2018learning} & Airplane & 42.17 & 13.84 & 3.819 & 1.810 & 77.59 & 98.52 & 6.188 \\
GCN-GAN~\citeyear{valsesia2019learning} & Airplane & 39.04 & 18.62 & 4.713 & 1.650 & 89.13 & 98.60 & 6.669 \\
TreeGAN~\citeyear{shu20193d} & Airplane & 39.37 & 8.40 & 4.323 & 1.953 & 83.86 & 99.67 & 15.646 \\
PointFlow~\citeyear{yang2019pointflow} & Airplane & 44.98 & 44.65 & 3.688 & 1.090 & 66.39 & 69.36 & 1.536 \\
ShapeGF~\citeyear{cai2020learning} & Airplane & 50.41 & 47.12 & 3.306 & 1.027 & 61.94 & 70.51 & 1.059 \\
DPM3D~\citeyear{luo2021diffusion} & Airplane & 48.71 & 45.47 & 3.276 & 1.061 & 64.83 & 75.12 & 1.067 \\
DDM (Ours) & Airplane & \textbf{50.79} & \textbf{48.23} & \textbf{3.223} & \textbf{1.021} & \textbf{60.98} & \textbf{69.12} & \textbf{1.044} \\ \hline
PC-GAM~\citeyear{achlioptas2018learning} & Chair & 46.23 & 22.14 & 13.436 & 3.104 & 69.67 & 100.00 & 6.649 \\
GCN-GAN~\citeyear{valsesia2019learning} & Chair & 39.84 & 35.09 & 15.354 & 2.213 & 77.86 & 95.80 & 21.708 \\
TreeGAN~\citeyear{shu20193d} & Chair & 38.02 & 6.77 & 14.936 & 3.613 & 74.92 & 100.00 & 13.282 \\
PointFlow~\citeyear{yang2019pointflow} & Chair & 41.86 & 43.38 & 13.631 & 1.856 & 66.13 & 68.40 & 12.474 \\
ShapeGF~\citeyear{cai2020learning} & Chair & 48.53 & 46.71 & 13.175 & 1.785 & \textbf{56.17} & \textbf{62.69} & \textbf{5.996} \\
DPM3D~\citeyear{luo2021diffusion} & Chair & 48.94 & 47.52 & 12.276 & 1.784 & 60.11 & 69.06 & 7.797 \\
DDM (Ours) & Chair & \textbf{49.32} & \textbf{47.96} & \textbf{12.192} & \textbf{1.773} & 57.36 & 63.62 & 6.117 \\ \hline
\end{tabular}
}
\caption{Comparison of pointclouds generation performance. CD is multiplied by $10^3$, EMD is multiplied by $10$ and JSD is multiplied by $10^2$.}
\label{tab: pcd generation}
\end{table*}

\subsection{Generation of High-Resolution Point Clouds}

To procure quantitative indicators, we integrate an additional point feature encoder and concatenate the feature vector with time embeddings, following the approach in~\cite{luo2021diffusion}. This facilitates conditional point cloud generation and allows us to apply most evaluation metrics for point cloud generation fairly to our methods. 

\paragraph{Experimental Setup.}
The experiment involves random sampling of point clouds from ShapeNet~\cite{chang2015shapenet}, a dataset comprising 51,127 shapes across 55 categories, as $\mathcal{X}^{(0)}$ at each iteration. The dataset is randomly partitioned into training, testing, and validation sets at ratios of $80\%, 15\%$, and $5\%$ respectively. A well-initialized $\mathcal{X}^{(T)}$ is obtained using a data-driven method with shapes from the training data. Given that point clouds only contain vertices $\mathbf{x}$, there are no shape regularizations,  \textit{i.e.}, $\lambda = \mathbf{0}, \lambda_{c} = 1.0$. We employ DDK for 500 steps in the diffusion process and set $i=50$ for equispaced sampling, implying that only the samples at steps $0, 50, 100, \cdots$ and $500$ are used for training. During the reverse diffusion sampling process, generating point clouds with the learned DDM takes only 10 steps.

 \begin{figure*}
    \centering
    \includegraphics[width=\linewidth]{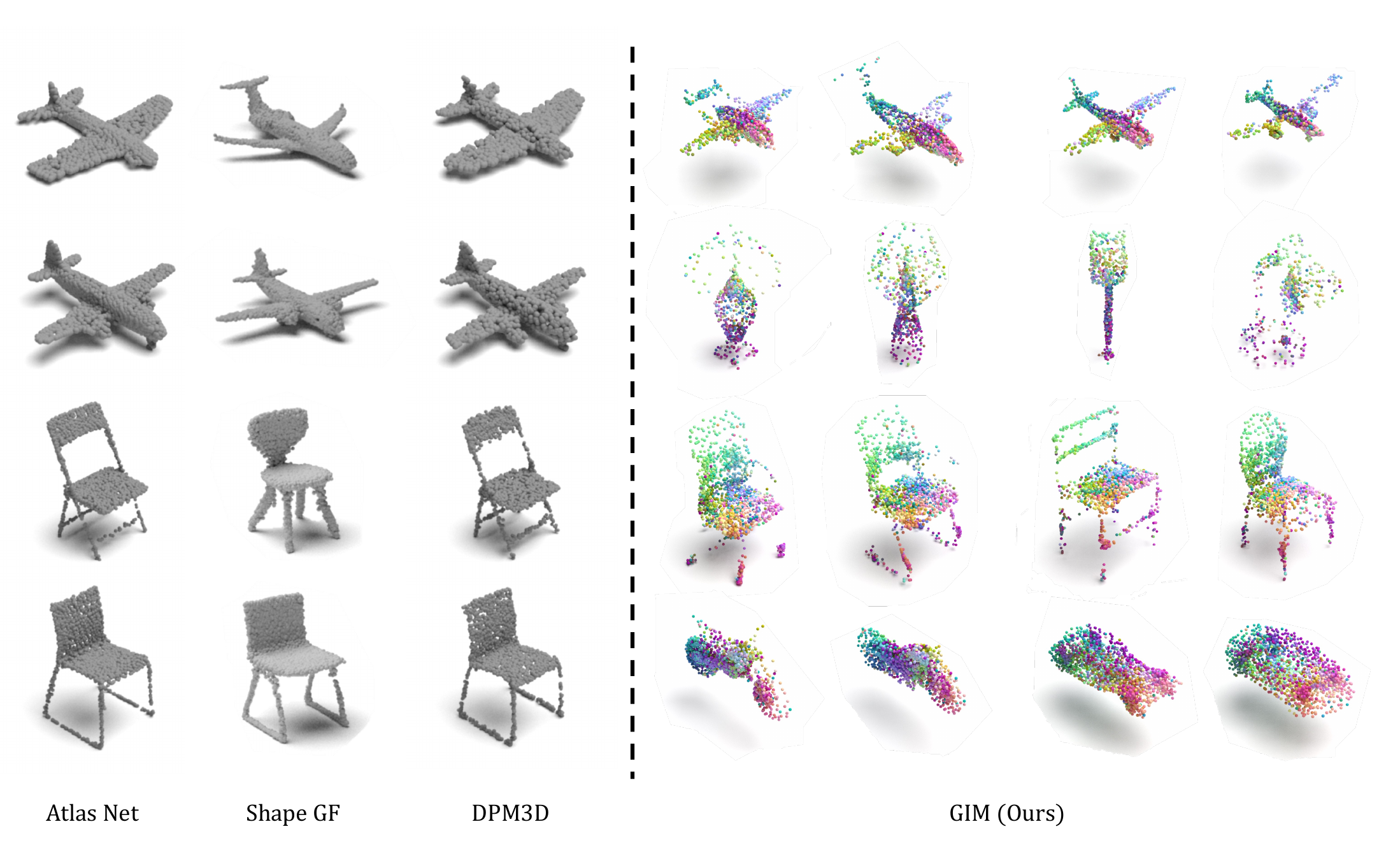}
    \caption{Comparison of different point cloud generation methods, namely, Atlas Net~\cite{groueix2018papier}, Shape GF~\cite{cai2020learning}, and DPM3D~\cite{luo2021diffusion}.}
    \label{fig: comparison pcl}
\end{figure*} 

\paragraph{Results and Discussion.}
As illustrated in Tab.~\ref{tab: pcd generation}, our proposed DDM exhibits exceptional performance across various evaluation metrics. Notably, DPM3D~\cite{luo2021diffusion} also treats point cloud generation as a diffusion process but employs GDK instead of DDK for the diffusion process. Despite DPM3D achieving competitive results, it necessitates numerous steps to sample a point cloud, whereas our model requires only 10 steps to generate a comparable shape, underscoring the advantages of DDK in 3D shape generation. Fig.~\ref{fig: comparison pcl} displays the point clouds generated by a simple DDM without a latent vector. The progressive generation results are depicted in Fig.~\ref{fig: progressive all}.

\begin{table}[]
    \centering
    \resizebox{\linewidth}{!}{
        \begin{tabular}{l|ccccccc}
         & \multicolumn{2}{c}{COV(\%,$\uparrow$)} & \multicolumn{2}{c}{MMD~($\downarrow$)} & \multicolumn{2}{c}{1-NNA~(\%,$\downarrow$)} & JSD~($\downarrow$) \\ \cline{2-8} 
                                     & CD    & EMD   & CD    & EMD   & CD    & EMD   & -     \\ \hline
        GDK & 49.25 & 47.51 & 3.416 & 1.110 & 62.59 & 71.48 & 1.096 \\
        DDK & 51.23 & 49.12 & 3.189 & 1.034 & 60.01 & 70.19 & 1.045
        \end{tabular}
    }
    \caption{Performance in latent space.}
    \label{tab:latent}
    \vspace{-12pt}
\end{table}

\paragraph{Extension to Latent Diffusion Models.} 
Latent diffusion models have been demonstrated to lessen the computational load during the training of diffusion models. Recently, LION~\cite{zeng2022lion} also applied this approach to point cloud generation. In this study, we apply DDK in the latent space by initially deforming each shape with DDK, followed by feeding them into the shape encoder network to obtain the latent code. As shown in Tab.~\ref{tab:latent}, DDK can enhance the model's performance in the latent space.
\subsection{Generation of Well-Manifold Meshes}
\begin{figure}
    \centering
    \includegraphics[width=\linewidth]{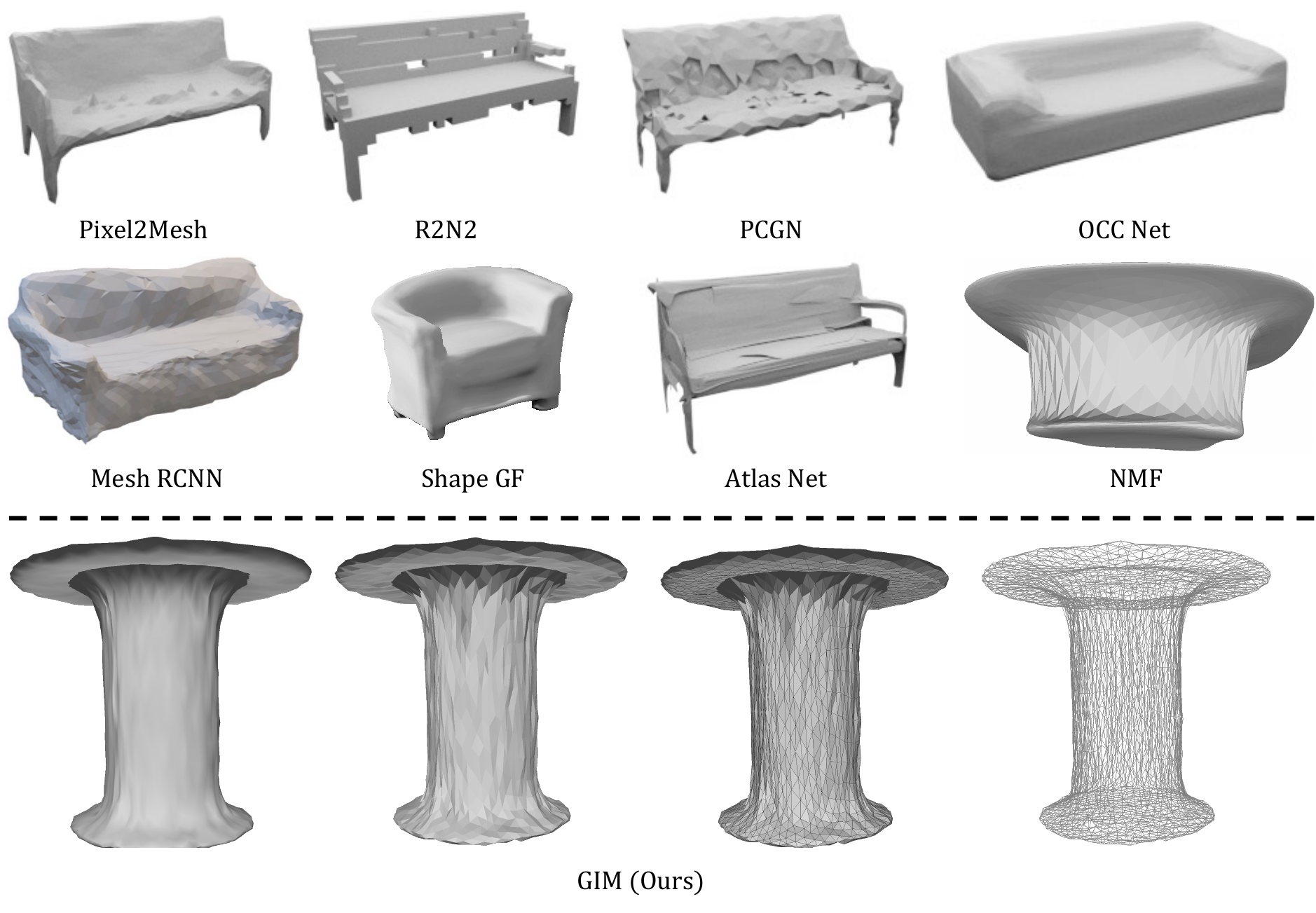}
    \caption{Comparative analysis of various mesh generation methodologies, namely, Pixel2Mesh~\cite{wang2018pixel2mesh}, R2N2~\cite{choy20163d}, PCGN~\cite{fan2017point}, OCC Net~\cite{mescheder2019occupancy}, Mesh RCNN~\cite{gkioxari2019mesh}, Shape GF~\cite{cai2020learning}, Atlas Net~\cite{groueix2018papier}, and NMF~\cite{gupta2020neural}.}
    \label{fig: comparison mesh}
    \vspace{-12pt}
\end{figure}

The task of generating meshes is significantly challenging due to the necessity of maintaining the correct topology amongst each vertex. A well-manifold mesh structure is crucial for applications such as graphic rendering. In our experiment, we utilized shapes from ShapeNet~\cite{chang2015shapenet} and a unit sphere with nearly 5,000 vertices as $\mathcal{X}^{(T)}$. For each sampled $\mathcal{X}^{(0)}$, we employed DDK for 2,000 steps during the diffusion process, setting $\lambda_{c}=1.0, \lambda_{e}=0.8, \lambda_{n}=0.01, \lambda_{l}=0.15, \lambda_{p}=0.01$. We established $i=50$ for equispaced sampling, necessitating 40 steps to sample a mesh.
\begin{table*}[!h]
\centering

\begin{tabular}{l|l|ccccccc}
\multirow{2}{*}{Init. of $\mathcal{X}^{(T)}$} &
  \multirow{2}{*}{Steps} &
  \multicolumn{2}{c}{COV(\%,$\uparrow$)} &
  \multicolumn{2}{c}{MMD~($\downarrow$)} &
  \multicolumn{2}{c}{1-NNA~(\%,$\downarrow$)} &
  JSD~($\downarrow$) \\ \cline{3-9} 
               &      & CD    & EMD   & CD    & EMD   & CD    & EMD   & -     \\ \hline
Gaussian noise & 1000 & 50.24 & 47.14 & 3.543 & 1.211 & 62.98 & 71.13 & 1.245 \\
Unit sphere    & 1000 & 50.37 & 47.42 & 3.452 & 1.179 & 61.31 & 69.81 & 1.104 \\
Average shape     & 500  & 50.79 & 48.23 & 3.223 & 1.021 & 60.98 & 69.12 & 1.044
\end{tabular}
\caption{The impact of different initializations of $\mathcal{X}^{T}$.}
\label{tab: ablation}
\end{table*}
As depicted in Fig.~\ref{fig: comparison mesh}, DDM can generate meshes that are comparable to those produced by other state-of-the-art methods in terms of visualization. However, due to the limited resolution of $\mathcal{X}^{(T)}$, it is currently unfeasible to generate meshes with complex geometric structures. Future work will focus on developing a superior $\mathcal{X}^{(T)}$ to enhance the geometric imitation model's representation capability for mesh generation. Fig.~\ref{fig: progressive all} illustrates the process of gradually deforming a unit sphere into target shapes using the geometric imitation model.

\paragraph{User Study of Generated Meshes.}
\begin{table*}[!h]
\centering
\begin{tabular}{ccccccccc}
  Pixel2Mesh & R2N2 & PCGN & OCCNet & MeshRCNN & ShapeGF & AtlasNet & NMF & DDM \\ \hline
    8\%       &  10\%    &   4\%   &   8\%     &   8\%       &   14\%      &    8\%      & 18\%    &  \textbf{22\%}
\end{tabular}
\caption{Comparison of the first rank ratio for various models.}
\label{tab: user study}
\end{table*}
In our user study, we utilized various methods to generate ten sets of mesh models with distinct themes. We then requested five users to select the best model from each set. Tab.~\ref{tab: user study} displays the ratio for each method that was selected as the best-quality model. The user study indicates that our generated meshes are preferred by a majority of users.

%\subsection{Ablation Study on $\mathcal{X}^{(T)}$}
\subsection{Ablation Study on $X^{(T)}$}

Tab.~\ref{tab: ablation} demonstrates the effect of different initialization methods on $\mathcal{X}^{T}$, suggesting that DDK displays robust generalization performance across diverse initialization methods. Nevertheless, a well-initialized $\mathcal{X}^{T}$ can significantly decrease the number of steps required for DDK to converge to $\mathcal{X}^{0}$.

\subsection{Broad Applications}

\begin{figure}
    \centering
    \includegraphics[width=\linewidth]{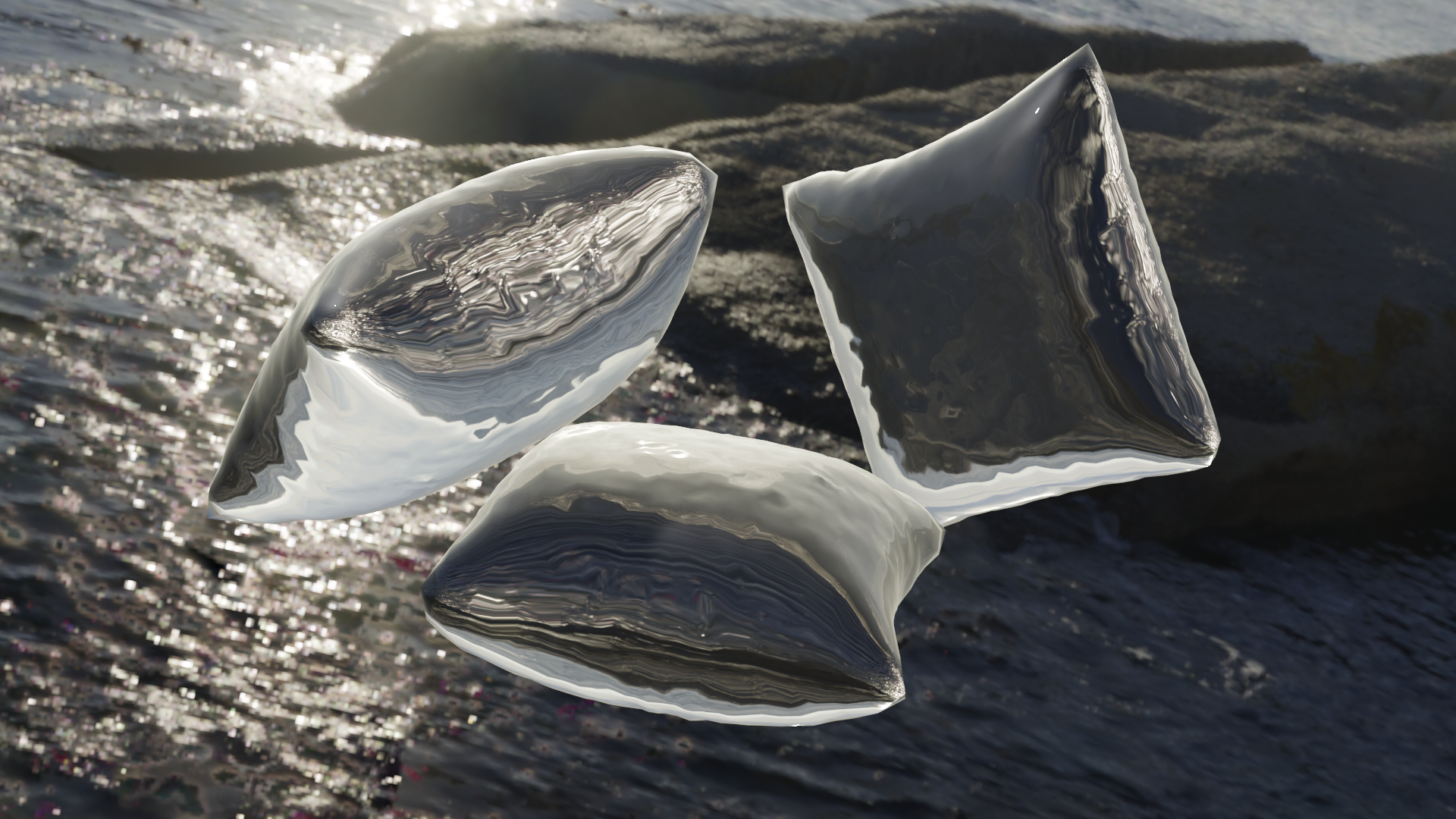}
    \caption{High-fidelity image rendered with the generated pillow mesh.}
    \label{fig: rendering}
\end{figure}

\paragraph{High-Fidelity Rendering.}
As illustrated in Fig.~\ref{fig: rendering}, we render high-fidelity images using Blender~\footnote{https://www.blender.org} without any post-processing. Owing to the well-manifold structure of the generated mesh, light can naturally penetrate the object (e.g., pillow), leading to realistic reflections in the rendered images.

\begin{figure*}[htb]
    \centering
    \includegraphics[width=\linewidth]{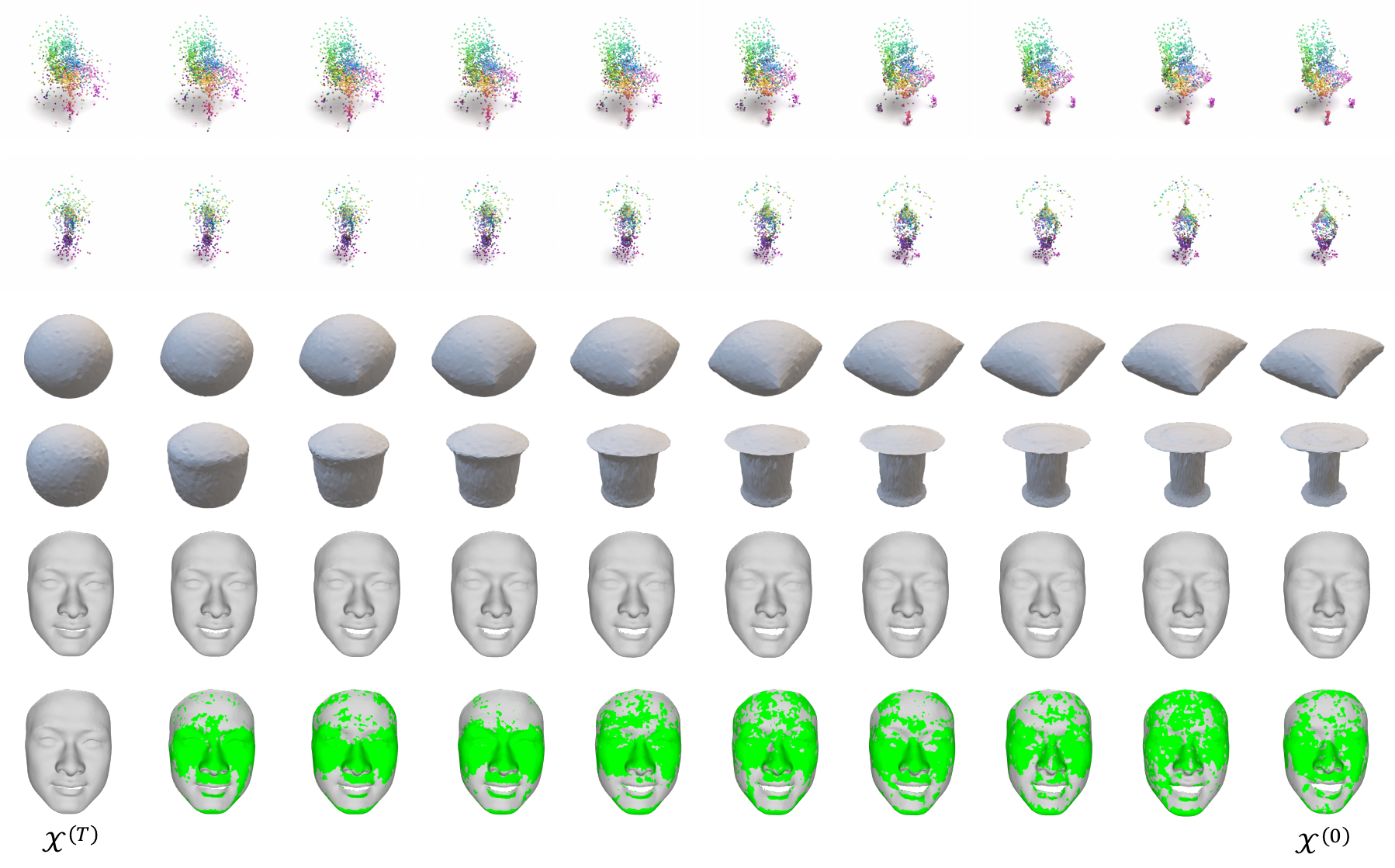}
    \caption{
    The first two rows and the next two rows represent the progressive generation processes for point clouds and meshes, respectively. The last two rows display the gradual animation of facial expressions, with the second to last row showing the original generated facial meshes. 
    $\mathcal{X}^{(T)}$ symbolizes the initial shape devoid of any expressions. To make comparisons easier, we highlight the differences between two consecutive facial meshes in the final row.
     }
    \label{fig: progressive all}
\end{figure*}

\paragraph{Facial Expression Animation}
As previously stated, DDK can capture fine-grained geometric structures. Consequently, we can modify a geometric imitation model to drive a template facial mesh to natural facial meshes with expressive features, as shown in Fig.~\ref{fig: progressive all}. To the best of our knowledge, we are the first to drive a facial mesh without any landmarks. Unlike previous generation tasks, we adapted a template facial mesh provided by ~\cite{zhang2004spacetime} as $\mathcal{X}^{(T)}$. We adopted a smaller step size $\eta$ and did not adopt equispaced sampling to avoid disrupting the facial topologies. The series of facial meshes that we generated fully demonstrates the potential of geometric imitation models in a variety of future applications.
\section{Conclusion}

In this study, we introduced a novel approach for the generation of superior quality point clouds and meshes. Our methodology leverages the recently developed Differential Deformation Kernel (DDK), which facilitates diffusion in a geometrically aware manner. Additionally, we utilize the Geometric Imitation Model (DDM), an innovative technique capable of reversing the diffusion process of three-dimensional shapes. We posit that our methodology has potential for extension to shape generation from images or textual data, presenting significant implications for diverse fields including robotics, gaming, and interactive design.

{
    \small
    \bibliographystyle{ieeenat_fullname}
    \bibliography{egbib}

\begin{thebibliography}{43}
\providecommand{\natexlab}[1]{#1}
\providecommand{\url}[1]{\texttt{#1}}
\expandafter\ifx\csname urlstyle\endcsname\relax
  \providecommand{\doi}[1]{doi: #1}\else
  \providecommand{\doi}{doi: \begingroup \urlstyle{rm}\Url}\fi

\bibitem[Achlioptas et~al.(2018)Achlioptas, Diamanti, Mitliagkas, and
  Guibas]{achlioptas2018learning}
Panos Achlioptas, Olga Diamanti, Ioannis Mitliagkas, and Leonidas Guibas.
\newblock Learning representations and generative models for 3d point clouds.
\newblock In \emph{International conference on machine learning}, pages 40--49.
  PMLR, 2018.

\bibitem[Cai et~al.(2020)Cai, Yang, Averbuch-Elor, Hao, Belongie, Snavely, and
  Hariharan]{cai2020learning}
Ruojin Cai, Guandao Yang, Hadar Averbuch-Elor, Zekun Hao, Serge Belongie, Noah
  Snavely, and Bharath Hariharan.
\newblock Learning gradient fields for shape generation.
\newblock In \emph{European Conference on Computer Vision}, pages 364--381.
  Springer, 2020.

\bibitem[Chang et~al.(2015)Chang, Funkhouser, Guibas, Hanrahan, Huang, Li,
  Savarese, Savva, Song, Su, et~al.]{chang2015shapenet}
Angel~X Chang, Thomas Funkhouser, Leonidas Guibas, Pat Hanrahan, Qixing Huang,
  Zimo Li, Silvio Savarese, Manolis Savva, Shuran Song, Hao Su, et~al.
\newblock Shapenet: An information-rich 3d model repository.
\newblock \emph{arXiv preprint arXiv:1512.03012}, 2015.

\bibitem[Chen et~al.(2018)Chen, Rubanova, Bettencourt, and
  Duvenaud]{chen2018neural}
Ricky~TQ Chen, Yulia Rubanova, Jesse Bettencourt, and David~K Duvenaud.
\newblock Neural ordinary differential equations.
\newblock \emph{Advances in neural information processing systems}, 31, 2018.

\bibitem[Chen and Zhang(2019)]{chen2019learning}
Zhiqin Chen and Hao Zhang.
\newblock Learning implicit fields for generative shape modeling.
\newblock In \emph{Proceedings of the IEEE/CVF Conference on Computer Vision
  and Pattern Recognition}, pages 5939--5948, 2019.

\bibitem[Choy et~al.(2016)Choy, Xu, Gwak, Chen, and Savarese]{choy20163d}
Christopher~B Choy, Danfei Xu, JunYoung Gwak, Kevin Chen, and Silvio Savarese.
\newblock 3d-r2n2: A unified approach for single and multi-view 3d object
  reconstruction.
\newblock In \emph{European conference on computer vision}, pages 628--644.
  Springer, 2016.

\bibitem[Dhariwal and Nichol(2021)]{dhariwal2021diffusion}
Prafulla Dhariwal and Alexander Nichol.
\newblock Diffusion models beat gans on image synthesis.
\newblock \emph{Advances in Neural Information Processing Systems},
  34:\penalty0 8780--8794, 2021.

\bibitem[Fan et~al.(2017)Fan, Su, and Guibas]{fan2017point}
Haoqiang Fan, Hao Su, and Leonidas~J Guibas.
\newblock A point set generation network for 3d object reconstruction from a
  single image.
\newblock In \emph{Proceedings of the IEEE conference on computer vision and
  pattern recognition}, pages 605--613, 2017.

\bibitem[Gadelha et~al.(2018)Gadelha, Wang, and
  Maji]{gadelha2018multiresolution}
Matheus Gadelha, Rui Wang, and Subhransu Maji.
\newblock Multiresolution tree networks for 3d point cloud processing.
\newblock In \emph{Proceedings of the European Conference on Computer Vision
  (ECCV)}, pages 103--118, 2018.

\bibitem[Gkioxari et~al.(2019)Gkioxari, Malik, and Johnson]{gkioxari2019mesh}
Georgia Gkioxari, Jitendra Malik, and Justin Johnson.
\newblock Mesh r-cnn.
\newblock In \emph{Proceedings of the IEEE/CVF International Conference on
  Computer Vision}, pages 9785--9795, 2019.

\bibitem[Goodfellow et~al.(2020)Goodfellow, Pouget-Abadie, Mirza, Xu,
  Warde-Farley, Ozair, Courville, and Bengio]{goodfellow2020generative}
Ian Goodfellow, Jean Pouget-Abadie, Mehdi Mirza, Bing Xu, David Warde-Farley,
  Sherjil Ozair, Aaron Courville, and Yoshua Bengio.
\newblock Generative adversarial networks.
\newblock \emph{Communications of the ACM}, 63\penalty0 (11):\penalty0
  139--144, 2020.

\bibitem[Grathwohl et~al.(2018)Grathwohl, Chen, Bettencourt, Sutskever, and
  Duvenaud]{grathwohl2018ffjord}
Will Grathwohl, Ricky~TQ Chen, Jesse Bettencourt, Ilya Sutskever, and David
  Duvenaud.
\newblock Ffjord: Free-form continuous dynamics for scalable reversible
  generative models.
\newblock \emph{arXiv preprint arXiv:1810.01367}, 2018.

\bibitem[Groueix et~al.(2018)Groueix, Fisher, Kim, Russell, and
  Aubry]{groueix2018papier}
Thibault Groueix, Matthew Fisher, Vladimir~G Kim, Bryan~C Russell, and Mathieu
  Aubry.
\newblock A papier-m{\^a}ch{\'e} approach to learning 3d surface generation.
\newblock In \emph{Proceedings of the IEEE conference on computer vision and
  pattern recognition}, pages 216--224, 2018.

\bibitem[Guo et~al.(2020)Guo, Wang, Hu, Liu, Liu, and Bennamoun]{guo2020deep}
Yulan Guo, Hanyun Wang, Qingyong Hu, Hao Liu, Li Liu, and Mohammed Bennamoun.
\newblock Deep learning for 3d point clouds: A survey.
\newblock \emph{IEEE transactions on pattern analysis and machine
  intelligence}, 43\penalty0 (12):\penalty0 4338--4364, 2020.

\bibitem[Gupta(2020)]{gupta2020neural}
Kunal Gupta.
\newblock \emph{Neural mesh flow: 3d manifold mesh generation via diffeomorphic
  flows}.
\newblock University of California, San Diego, 2020.

\bibitem[Ho et~al.(2020)Ho, Jain, and Abbeel]{ho2020denoising}
Jonathan Ho, Ajay Jain, and Pieter Abbeel.
\newblock Denoising diffusion probabilistic models.
\newblock \emph{Advances in Neural Information Processing Systems},
  33:\penalty0 6840--6851, 2020.

\bibitem[Jarzynski(1997)]{jarzynski1997equilibrium}
Christopher Jarzynski.
\newblock Equilibrium free-energy differences from nonequilibrium measurements:
  A master-equation approach.
\newblock \emph{Physical Review E}, 56\penalty0 (5):\penalty0 5018, 1997.

\bibitem[Kingma and Welling(2013)]{kingma2013auto}
Diederik~P Kingma and Max Welling.
\newblock Auto-encoding variational bayes.
\newblock \emph{arXiv preprint arXiv:1312.6114}, 2013.

\bibitem[Klokov et~al.(2020)Klokov, Boyer, and Verbeek]{klokov2020discrete}
Roman Klokov, Edmond Boyer, and Jakob Verbeek.
\newblock Discrete point flow networks for efficient point cloud generation.
\newblock In \emph{European Conference on Computer Vision}, pages 694--710.
  Springer, 2020.

\bibitem[Kolotouros et~al.(2019)Kolotouros, Pavlakos, and
  Daniilidis]{kolotouros2019convolutional}
Nikos Kolotouros, Georgios Pavlakos, and Kostas Daniilidis.
\newblock Convolutional mesh regression for single-image human shape
  reconstruction.
\newblock In \emph{Proceedings of the IEEE/CVF Conference on Computer Vision
  and Pattern Recognition}, pages 4501--4510, 2019.

\bibitem[Liu et~al.(2023)Liu, Park, Azadi, Zhang, Chopikyan, Hu, Shi, Rohrbach,
  and Darrell]{liu2023more}
Xihui Liu, Dong~Huk Park, Samaneh Azadi, Gong Zhang, Arman Chopikyan, Yuxiao
  Hu, Humphrey Shi, Anna Rohrbach, and Trevor Darrell.
\newblock More control for free! image synthesis with semantic diffusion
  guidance.
\newblock In \emph{Proceedings of the IEEE/CVF Winter Conference on
  Applications of Computer Vision}, pages 289--299, 2023.

\bibitem[Luo and Hu(2021)]{luo2021diffusion}
Shitong Luo and Wei Hu.
\newblock Diffusion probabilistic models for 3d point cloud generation.
\newblock In \emph{Proceedings of the IEEE/CVF Conference on Computer Vision
  and Pattern Recognition}, pages 2837--2845, 2021.

\bibitem[Mescheder et~al.(2019)Mescheder, Oechsle, Niemeyer, Nowozin, and
  Geiger]{mescheder2019occupancy}
Lars Mescheder, Michael Oechsle, Michael Niemeyer, Sebastian Nowozin, and
  Andreas Geiger.
\newblock Occupancy networks: Learning 3d reconstruction in function space.
\newblock In \emph{Proceedings of the IEEE/CVF conference on computer vision
  and pattern recognition}, pages 4460--4470, 2019.

\bibitem[Park et~al.(2019)Park, Florence, Straub, Newcombe, and
  Lovegrove]{park2019deepsdf}
Jeong~Joon Park, Peter Florence, Julian Straub, Richard Newcombe, and Steven
  Lovegrove.
\newblock Deepsdf: Learning continuous signed distance functions for shape
  representation.
\newblock In \emph{Proceedings of the IEEE/CVF conference on computer vision
  and pattern recognition}, pages 165--174, 2019.

\bibitem[Qi et~al.(2017)Qi, Yi, Su, and Guibas]{qi2017pointnet++}
Charles~Ruizhongtai Qi, Li Yi, Hao Su, and Leonidas~J Guibas.
\newblock Pointnet++: Deep hierarchical feature learning on point sets in a
  metric space.
\newblock \emph{Advances in neural information processing systems}, 30, 2017.

\bibitem[Shu et~al.(2019)Shu, Park, and Kwon]{shu20193d}
Dong~Wook Shu, Sung~Woo Park, and Junseok Kwon.
\newblock 3d point cloud generative adversarial network based on tree
  structured graph convolutions.
\newblock In \emph{Proceedings of the IEEE/CVF international conference on
  computer vision}, pages 3859--3868, 2019.

\bibitem[Sohl-Dickstein et~al.(2015)Sohl-Dickstein, Weiss, Maheswaranathan, and
  Ganguli]{sohl2015deep}
Jascha Sohl-Dickstein, Eric Weiss, Niru Maheswaranathan, and Surya Ganguli.
\newblock Deep unsupervised learning using nonequilibrium thermodynamics.
\newblock In \emph{International Conference on Machine Learning}, pages
  2256--2265. PMLR, 2015.

\bibitem[Song et~al.(2020)Song, Meng, and Ermon]{song2020denoising}
Jiaming Song, Chenlin Meng, and Stefano Ermon.
\newblock Denoising diffusion implicit models.
\newblock \emph{arXiv preprint arXiv:2010.02502}, 2020.

\bibitem[Suchde et~al.(2022)Suchde, Jacquemin, and Davydov]{suchde2022point}
Pratik Suchde, Thibault Jacquemin, and Oleg Davydov.
\newblock Point cloud generation for meshfree methods: An overview.
\newblock \emph{Archives of Computational Methods in Engineering}, pages 1--27,
  2022.

\bibitem[Sun et~al.(2020)Sun, Wang, Liu, Siegel, and Sarma]{sun2020pointgrow}
Yongbin Sun, Yue Wang, Ziwei Liu, Joshua Siegel, and Sanjay Sarma.
\newblock Pointgrow: Autoregressively learned point cloud generation with
  self-attention.
\newblock In \emph{Proceedings of the IEEE/CVF Winter Conference on
  Applications of Computer Vision}, pages 61--70, 2020.

\bibitem[Valsesia et~al.(2018)Valsesia, Fracastoro, and
  Magli]{valsesia2018learning}
Diego Valsesia, Giulia Fracastoro, and Enrico Magli.
\newblock Learning localized generative models for 3d point clouds via graph
  convolution.
\newblock In \emph{International conference on learning representations}, 2018.

\bibitem[Valsesia et~al.(2019)Valsesia, Fracastoro, and
  Magli]{valsesia2019learning}
Diego Valsesia, Giulia Fracastoro, and Enrico Magli.
\newblock Learning localized generative models for 3d point clouds via graph
  convolution.
\newblock In \emph{International conference on learning representations}, 2019.

\bibitem[Wang et~al.(2018)Wang, Zhang, Li, Fu, Liu, and
  Jiang]{wang2018pixel2mesh}
Nanyang Wang, Yinda Zhang, Zhuwen Li, Yanwei Fu, Wei Liu, and Yu-Gang Jiang.
\newblock Pixel2mesh: Generating 3d mesh models from single rgb images.
\newblock In \emph{Proceedings of the European conference on computer vision
  (ECCV)}, pages 52--67, 2018.

\bibitem[Wu et~al.(2018)Wu, Zhang, Zhang, Zhang, Freeman, and
  Tenenbaum]{wu2018learning}
Jiajun Wu, Chengkai Zhang, Xiuming Zhang, Zhoutong Zhang, William~T Freeman,
  and Joshua~B Tenenbaum.
\newblock Learning shape priors for single-view 3d completion and
  reconstruction.
\newblock In \emph{Proceedings of the European Conference on Computer Vision
  (ECCV)}, pages 646--662, 2018.

\bibitem[Wu et~al.(2015)Wu, Song, Khosla, Yu, Zhang, Tang, and Xiao]{wu20153d}
Zhirong Wu, Shuran Song, Aditya Khosla, Fisher Yu, Linguang Zhang, Xiaoou Tang,
  and Jianxiong Xiao.
\newblock 3d shapenets: A deep representation for volumetric shapes.
\newblock In \emph{Proceedings of the IEEE conference on computer vision and
  pattern recognition}, pages 1912--1920, 2015.

\bibitem[Xu et~al.(2021)Xu, Tong, and Stilla]{xu2021voxel}
Yusheng Xu, Xiaohua Tong, and Uwe Stilla.
\newblock Voxel-based representation of 3d point clouds: Methods, applications,
  and its potential use in the construction industry.
\newblock \emph{Automation in Construction}, 126:\penalty0 103675, 2021.

\bibitem[Yang et~al.(2019)Yang, Huang, Hao, Liu, Belongie, and
  Hariharan]{yang2019pointflow}
Guandao Yang, Xun Huang, Zekun Hao, Ming-Yu Liu, Serge Belongie, and Bharath
  Hariharan.
\newblock Pointflow: 3d point cloud generation with continuous normalizing
  flows.
\newblock In \emph{Proceedings of the IEEE/CVF International Conference on
  Computer Vision}, pages 4541--4550, 2019.

\bibitem[Yang et~al.(2018)Yang, Feng, Shen, and Tian]{yang2018foldingnet}
Yaoqing Yang, Chen Feng, Yiru Shen, and Dong Tian.
\newblock Foldingnet: Point cloud auto-encoder via deep grid deformation.
\newblock In \emph{Proceedings of the IEEE conference on computer vision and
  pattern recognition}, pages 206--215, 2018.

\bibitem[Zeng et~al.(2022)Zeng, Vahdat, Williams, Gojcic, Litany, Fidler, and
  Kreis]{zeng2022lion}
Xiaohui Zeng, Arash Vahdat, Francis Williams, Zan Gojcic, Or Litany, Sanja
  Fidler, and Karsten Kreis.
\newblock Lion: Latent point diffusion models for 3d shape generation.
\newblock \emph{arXiv preprint arXiv:2210.06978}, 2022.

\bibitem[Zhang et~al.(2004)Zhang, Snavely, Curless, and
  Seitz]{zhang2004spacetime}
Li Zhang, Noah Snavely, Brian Curless, and Steven~M Seitz.
\newblock Spacetime faces: high resolution capture for modeling and animation.
\newblock In \emph{ACM SIGGRAPH 2004 Papers}, pages 548--558. 2004.

\bibitem[Zhang et~al.(2018)Zhang, Zhang, Zhang, Tenenbaum, Freeman, and
  Wu]{zhang2018learning}
Xiuming Zhang, Zhoutong Zhang, Chengkai Zhang, Josh Tenenbaum, Bill Freeman,
  and Jiajun Wu.
\newblock Learning to reconstruct shapes from unseen classes.
\newblock \emph{Advances in neural information processing systems}, 31, 2018.

\bibitem[Zuffi et~al.(2017)Zuffi, Kanazawa, Jacobs, and Black]{zuffi20173d}
Silvia Zuffi, Angjoo Kanazawa, David~W Jacobs, and Michael~J Black.
\newblock 3d menagerie: Modeling the 3d shape and pose of animals.
\newblock In \emph{Proceedings of the IEEE conference on computer vision and
  pattern recognition}, pages 6365--6373, 2017.

\bibitem[Zuffi et~al.(2018)Zuffi, Kanazawa, and Black]{zuffi2018lions}
Silvia Zuffi, Angjoo Kanazawa, and Michael~J Black.
\newblock Lions and tigers and bears: Capturing non-rigid, 3d, articulated
  shape from images.
\newblock In \emph{Proceedings of the IEEE conference on Computer Vision and
  Pattern Recognition}, pages 3955--3963, 2018.

\end{thebibliography}
}

\clearpage
\section{Related Works}
\subsection{Generation of Point Clouds}
Initial methodologies for point cloud generation~\cite{achlioptas2018learning,gadelha2018multiresolution} transformed the problem into a matrix generation task, which facilitated the application of pre-existing generative models originally designed for image generation~\cite{valsesia2018learning,shu20193d,sun2020pointgrow,yang2019pointflow}. However, these methods~\cite{kingma2013auto,goodfellow2020generative} presented significant limitations, such as the inability to generate point clouds with varying numbers of points and a lack of permutation invariance—a property often crucial for effective point cloud generation in practical applications.

An alternative approach treated point clouds as samples drawn from a point distribution, which led to the exploration of likelihood-based methods for point cloud generation and modeling~\cite{klokov2020discrete,wu20153d,cai2020learning,chen2018neural,grathwohl2018ffjord}. These methods hold significant promise for advancing the field of 3D shape generation.

\subsection{Generation of Meshes}
In the realm of mesh generation leveraging machine learning techniques, the employed methodologies can be broadly classified into two primary categories: direct and indirect methods. 

\paragraph{Indirect Methods} These approaches involve predicting 3D geometry as a spatial distribution of voxels~\cite{wu2018learning,zhang2018learning}, point clouds~\cite{fan2017point,yang2018foldingnet}, or as an implicit function representing the signed distance from the surface. However, voxel and point cloud prediction methods often struggle to produce high-resolution outputs, leading to noisy or ineffective iso-surface extraction~\cite{groueix2018papier}. In implicit methods~\cite{mescheder2019occupancy,chen2019learning}, a neural network is supplied with a latent code and a query point to predict either the Truncated Signed Distance Function (TSDF)~\cite{park2019deepsdf} value or the binary occupancy of the point, but these methods are computationally intensive.

\paragraph{Direct Methods} Early approaches for direct mesh generation focused on predicting the parameters of category-based mesh models~\cite{zuffi2018lions,zuffi20173d,kolotouros2019convolutional}. However, these methods were limited to producing manifold meshes for object categories with pre-existing parameterized manifold meshes. Recent advancements in topological priors~\cite{wang2018pixel2mesh} have enhanced the ability to generate meshes for a broad array of categories, facilitating the successful generation of more complex and diverse meshes.

In summary, while indirect methods face challenges in output resolution and computational efficiency, direct methods have evolved to overcome category limitations, leading to more versatile and high-quality mesh generation.

\paragraph{Diffusion Probabilistic Models}
This study delves into a diffusion process closely associated with the Denoising Diffusion Probabilistic Model (DDPM), as outlined in previous work~\cite{ho2020denoising}. DDPM, an enhanced variant of the Diffusion Probabilistic Model (DPM), demonstrates superior capabilities in modeling the intricate distributions of natural images, thereby enabling the generation of unconditional images from noise inputs. Recent research~\cite{dhariwal2021diffusion, liu2023more} has suggested methods for controlling image synthesis using a reference signal, such as class, image, text, or embedding.
Expanding on this area, Luo et al.~\cite{luo2021diffusion} introduced DPM3D, an innovative probabilistic model capable of generating highly realistic point clouds.

Our methodology views point clouds as samples drawn from a specific distribution. However, rather than attempting to learn the posterior distribution of points, we focus on understanding the geometric distribution through our novel Differential Deformation Kernel (DDK). We employ a probabilistic geometric imitation model and utilize the imitation process to capture the complex deformations of 3D shapes. This results in training and sampling methodologies that differ significantly from previous approaches to traditional DPM.

\section{Point transformer network}

\begin{figure}
    \centering
    \includegraphics[width=\linewidth]{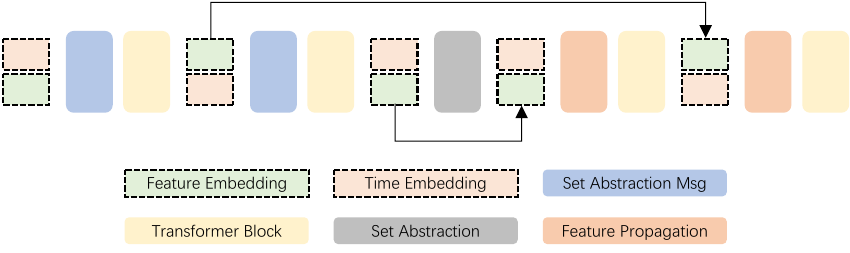}
    \caption{
    The architecture of the Point Transformer Network. The input vertices are initially processed through a Multilayer Perceptron (MLP) composed of two fully connected layers. The output features are also processed through an MLP with three linear layers to regress the offset for each vertex.
    }
    \label{fig: network structure}
\end{figure}
Recently, transformer blocks have shown a strong capability to extract features in natural language processes and computer vision. In this work, we use the PointNet++~\cite{qi2017pointnet++} as the foundational feature extractor framework. We incorporate a transformer block between two consecutive point blocks to enhance the ability to capture global features, as depicted in Fig.~\ref{fig: network structure}. Additionally, a time scale shifting module~\cite{ho2020denoising} is placed before the feature extraction layers. All other settings, such as the number of channels, are kept the same as in~\cite{qi2017pointnet++}.

It is important to note that we only feed the vertices to our network for both point clouds and meshes. The different geometric properties are injected through various shape regularizations.

\section{Experiments configuration}

We implement DDK mainly depending on the PyTorch3D library~\footnote{https://pytorch3d.org}, which is an efficient extension toolbox for PyTorch to deal with 3D shapes.
We conduct all the experiments on a single NVIDIA A100 GPU chip with 40GB of memory.
Comprehensive details of our experimental configurations are available in Tab.~\ref{tab: configuration details}.

\begin{table}[htb]
\caption{Configuration details.}
\centering
\label{tab: configuration details}
\resizebox{0.8\linewidth}{!}{
\begin{tabular}{lccc}
\multicolumn{1}{l|}{}             & pcl    & mesh   & face   \\ \hline
\multicolumn{4}{c}{Data-driven $\mathcal{X}^{(T)}$}          \\ \hline
\multicolumn{1}{l|}{npoints}      & 5k     & -      & -      \\
\multicolumn{1}{l|}{steps}        & 500    & -      & -      \\
\multicolumn{1}{l|}{$\lambda_c$}  & 1.0    & -      & -      \\
\multicolumn{1}{l|}{$\lambda_e$}  & 0.0    & -      & -      \\
\multicolumn{1}{l|}{$\lambda_n$}  & 0.0    & -      & -      \\
\multicolumn{1}{l|}{$\lambda_l$}  & 0.0    & -      & -      \\
\multicolumn{1}{l|}{$\lambda_p$}  & 0.01   & -      & -      \\
\multicolumn{1}{l|}{$\eta$}       & 1.0    & -      & -      \\ \hline
\multicolumn{4}{c}{Diffusion procedure with DDK}             \\ \hline
\multicolumn{1}{l|}{steps}        & 500    & 2k     & 500    \\
\multicolumn{1}{l|}{$\lambda_c$}  & 1.0    & 1.0    & 1.0    \\
\multicolumn{1}{l|}{$\lambda_e$}  & 0.0    & 0.8    & 0.8    \\
\multicolumn{1}{l|}{$\lambda_n$}  & 0.0    & 0.01   & 0.01   \\
\multicolumn{1}{l|}{$\lambda_l$}  & 0.0    & 0.15   & 0.15   \\
\multicolumn{1}{l|}{$\lambda_p$}  & 0.01   & 0.01   & 0.01   \\
\multicolumn{1}{l|}{$\eta$}       & 1.0    & 1.0    & 0.1    \\ \hline
\multicolumn{4}{c}{Geometric imitation learning}             \\
\multicolumn{1}{l|}{iterations} & 100k   & 100k   & 100k   \\
\multicolumn{1}{l|}{batch size}   & 32     & 32     & 32     \\
\multicolumn{1}{l|}{$\beta_t$}    & 0.05   & 0.05   & 0.01   \\
\multicolumn{1}{l|}{optimizer}    & AdamW  & AdamW  & AdamW  \\
\multicolumn{1}{l|}{lr}           & 2e-4   & 2e-4   & 2e-4   \\
\multicolumn{1}{l|}{weight decay} & 1e-6   & 1e-6   & 1e-6   \\
\multicolumn{1}{l|}{lr scheduler} & cosine & cosine & cosine
\end{tabular}
}
\end{table}
\begin{figure*}
    \centering
    \includegraphics[width=\linewidth]{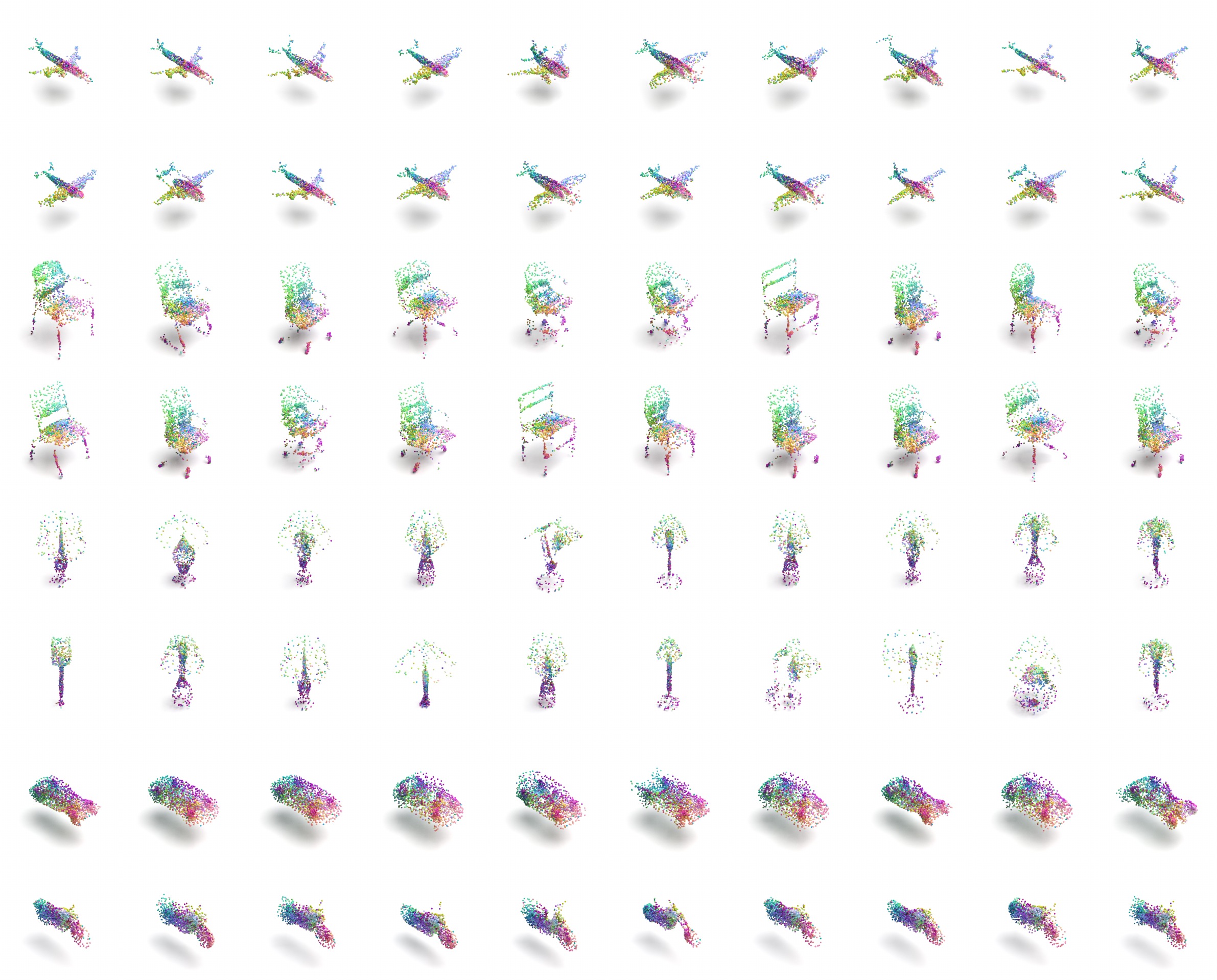}
    \caption{
    Additional examples of point cloud samples.
    }
    \label{fig: more pcl samples}
\end{figure*}
\section{Evaluation metrics for pointclouds.}
Following prior works, we use the Chamfer Distance (CD) and the Earth Mover's Distance (EMD) to assess the quality of reconstructed point clouds, following previous work~\cite{achlioptas2018learning}. 
For measuring the quality of the generated point clouds, we utilize the Minimum Matching Distance (MMD), Coverage Score (COV), 1-NN classifier accuracy (1-NNA), and Jenson-Shannon Divergence (JSD)~\cite{yang2019pointflow}. 
The MMD score evaluates the similarity between the generated samples and the reference samples, while the COV score detects potential mode collapse. 
The 1-NNA score is calculated by training a 1-NN classifier on the reference samples and testing it on both the reference and generated samples. The performance of the classifier is expected to be close to random guessing (i.e., 50\% accuracy) for well-generated samples. 
Finally, the JSD score is used to measure the disparity between the distributions of the generated and reference point sets.

\begin{figure}
    \centering
    \includegraphics[width=\linewidth]{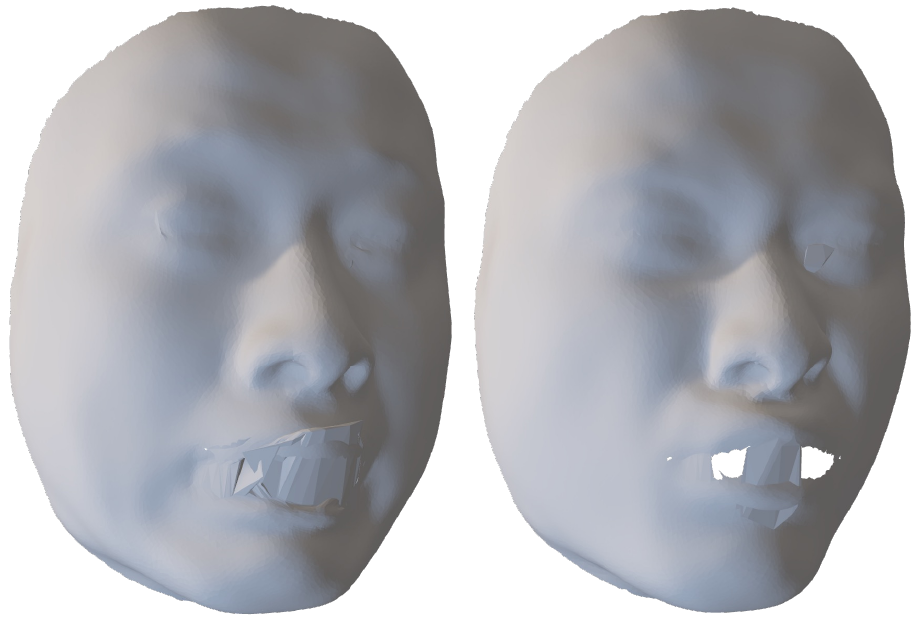}
    \caption{
    Examples of face generation failures.
    }
    \label{fig: face failure case}
\end{figure}
\begin{figure}
    \centering
    \includegraphics[width=\linewidth]{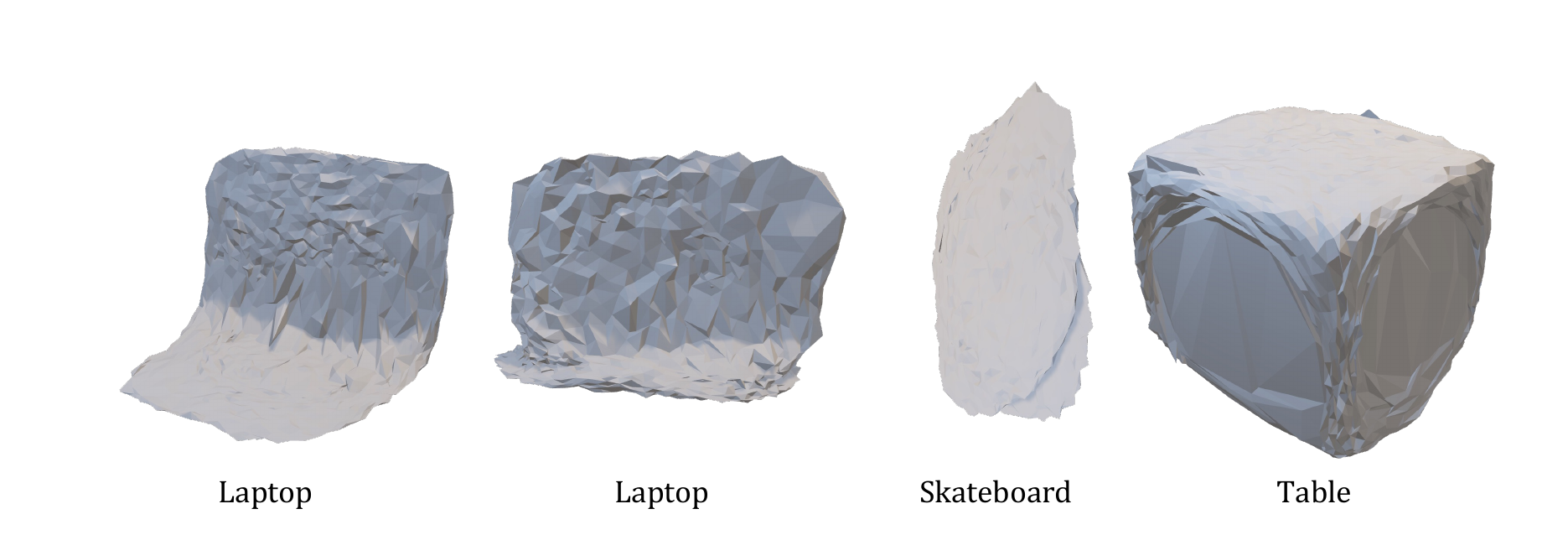}
    \caption{
    Examples of mesh generation failures.
    }
    \label{fig: mesh failure case}
\end{figure}

\section{More results}

\begin{figure}
    \centering
    \includegraphics[width=\linewidth]{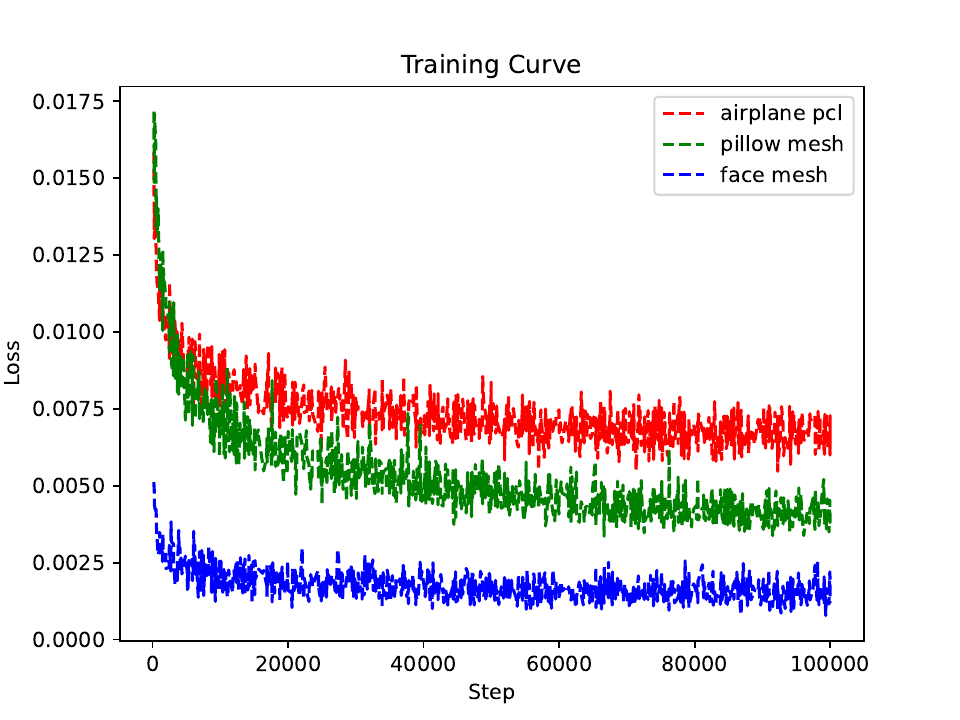}
    \caption{Curve of training loss.}
    \label{fig: training loss curve}
\end{figure}

In Fig.~\ref{fig: training loss curve}, we plot the loss curve during the geometric imitation learning procedure.

In Fig.~\ref{fig: more pcl samples}, we visualize more pointclouds generated via DDM.

\section{Failure cases and limitations}

Due to the limited adaptation of a unit sphere mesh as $\mathcal{X}^{T}$, it presents a challenge for DDM to effectively generate meshes with cohesive geometric structures. 
As illustrated in Fig.~\ref{fig: mesh failure case}, certain meshes have proven difficult to deform, particularly those with intricate details.
While we employ facial template adaptation for animation, we have discovered that having a well-structured topology for our template is crucial for achieving favorable mesh outcomes during deformation. 
As indicated in Fig.~\ref{fig: face failure case}, DDM is susceptible to yielding a disordered local structure surrounding the mouth area.

\end{document}